\documentclass{article}
\usepackage[dvips]{graphics}
\usepackage{amssymb}
\begin{document}

\title{The Forest Method as a New Parallel Tree Method
with the Sectional Voronoi Tessellation}
\author{Hideki Yahagi, Masao Mori, and Yuzuru Yoshii\footnote{Also at the Research Center for the Early Universe, University of Tokyo, Tokyo 113-0033, Japan}\\ {\it Institute of Astronomy, University of Tokyo, Mitaka, Tokyo 181-8588, Japan}}
\date{}
\maketitle
\begin{abstract}
We have developed a new parallel tree method which will be called the forest method hereafter.  This new method uses the sectional Voronoi tessellation (SVT) for the domain decomposition.  The SVT decomposes a whole space into polyhedra and allows their flat borders to move by assigning different weights. The forest method determines these weights based on the load balancing among processors by means of the over-load diffusion (OLD). Moreover, since all the borders are flat, before receiving the data from other processors, each processor can collect enough data to calculate the gravity force with precision.  Both the SVT and the OLD are coded in a highly vectorizable manner to accommodate on vector parallel processors.  The parallel code based on the forest method with the Message Passing Interface is run on various platforms so that a wide portability is guaranteed.  Extensive calculations with 15 processors of Fujitsu VPP300/16R indicate that the code can calculate the gravity force exerted on $10^5$  particles in each second for some ideal dark halo. This code is found to enable an $N$-body simulation with $10^7$ or more particles for a wide dynamic range and is therefore a very powerful tool for the study of galaxy formation and large-scale structure in the universe.
$$$$
\noindent
{\it Subject headings:} ~~ methods: numerical --- galaxies: formation --- galaxies: kinematics and dynamics
\end{abstract} 

\section{Introduction}
An $N$-body method, which calculates the gravity force exerted on constituent
particles and traces each of their trajectories in the phase space, provides 
a powerful tool for the study of formation and evolution of galaxies 
consisting of stars and dark matter.  Among many algorithms proposed so far 
for the force calculation, the particle-particle (PP) method uses an exact 
sum of gravitational interactions between all pairs of $N$ particles in a 
system so that the computational time scales as $N^2$.  Unless otherwise 
performed on a special purpose computer (Sugimoto {\it et al.} 1990), 
$N$-body simulations based on the PP method are therefore impractical when 
the number of particles is beyond $10^4$.  More elaborate algorithms have 
been developed, and those achieving much less time complexity while keeping 
the acceleration errors small mainly include the particle-mesh (PM) method 
(e.g. Miller 1978), the tree method (Appel 1985; Barnes \& Hut 1986; 
Jernigan \& Porter 1989) and the fast multipole method (FMM)
(Greengard \& Rokhlin 1987; Anderson 1992). 

The PM method determines the gravitational potential field from the discrete 
density field by solving the Poisson equation with the fast Fourier transform.
The computational time of the PM method scales as $N \log N$, where $N$ 
represents the number of the meshes and is usually taken to be of the 
same order with the number of particles.  The gravity forces from distant 
particles are calculated with high accuracy, but those from nearby particles 
are strongly softened.  Since the short-range two-body relaxation for nearby 
particles is suppressed accordingly, 
the PM method is favorable when the number of particles in the simulation 
is much smaller than in the real system.  This merit however competes with 
the corresponding resolution of the simulation 
limited by the grid spacing adopted.  Attempts to improve
the competing situation have been made in the following modifications to 
the original PM method.  That is, the particle-particle particle-mesh 
(P$^3$M) method applies the PP algorithm to nearby particles and the PM 
algorithm only to distant particles (Hockney \& Eastwood 1988; Efstathiou 
{\it et al.} 1985), and the adaptive particle-mesh (APM) method introduces 
more meshes preferentially to the high density regions (Villumsen 1989).  
A third alternative is a hybrid of these two modifications (Couchman 1991). 

The tree method on the other hand uses a similar concept to the PP method 
but decreases the number of gravitational interactions per particle from 
$O(N)$ to $O(\log N)$.  Thereby the computational time is contrived to scale 
as $N \log N$ like the PM method.  This decrease is achieved by grouping 
distant particles into clusters hierarchically, and a tree is used as a data 
structure to represent the hierarchy of the clusters.  Then the so-called
interaction list consisting of particles and clusters is constructed for
each particle for which the force is to be calculated.  [When all particles 
are included individually in this list, the tree method becomes equivalent 
to the PP method.]  A practical way of constructing the list is as follows:  
First, the root cluster of the tree is added to the temporary list.  If the cluster
added to the list does not meet the opening criterion for the required error level,
the cluster is removed from the list, and the clusters and particles of which
the removed cluster is composed at the next level is added to the list. 
This process of modifying the list, which is called the cell opening, is iterated until
all the clusters in the temporary list meet the opening criterion.
We here note that the FMM, although similar to the tree method, 
includes not only particle-particle and particle-cluster interactions, but 
also cluster-cluster interactions, so that the time complexity is further 
reduced (Greengard \& Rokhlin 1987; Anderson 1992). 

The tree method and the FMM, if paralleled, can handle more particles 
compared to the serial ones. The parallel computing is carried out by 
assigning separate processors to decomposed regions of the whole space.  
In order to maximize the computing performance, it is important to keep the 
data transfer time much shorter than the computational time, and this is 
achieved by making the surface of decomposed regions as small as possible.  
An intuitive example is a homogeneous configuration where the round and flat 
surface encloses a certain volume with the minimum area.

Different decomposition schemes are adopted in existing parallel tree
methods such as the orthogonal recursive bisection (ORB) method (Salmon 
1990; Dubinski 1996) and the parallel hashed oct-tree (HOT) method (Warren
\& Salmon 1993, 1995).  The ORB method decomposes a whole space into 
rectangular parallelepipeds by recursively bisecting regions perpendicular 
to an axis which is cyclically changed.  Such decomposed domains become 
elongated if particles are distributed in highly clustered manner.
In the HOT method, the multidimensional position data for the particles 
are mapped onto one dimensional integer array, called key, in such a way 
that the neighbor particles in the original space are mapped as neighbors
also in the key array.  Then, the space decomposition is equivalent to 
splitting the key array, and the decomposed domains are disconnected mostly
with indented borders.
In sharp contrast to the ORB and HOT methods, the forest method, which 
we have developed and will be described in this paper, uses the Voronoi 
decomposition scheme.  This method therefore ensures that the decomposed 
domains have flat borders and their elongation is suppressed, because 
the generator points are allowed to move as if the repulsive force operates
among them (Okabe {\it et al.} 1992).

In this paper we show
that a parallel tree code based on the Voronoi decomposition scheme realizes
$N$-body simulations with $10^7$ or more particles at a reasonable level 
of acceleration errors.  In \ref{frst:sect:tree} we review the basic features of the 
tree method which help to understand our strategy of newly constructing 
a faster parallel tree code or the forest code.  In \S \ref{frst:sect:frst} we describe the 
algorithm of the forest code which is developed with the Message Passing 
Interface (MPI).  Since the MPI is supported from the network of PCs to many 
types of supercomputers, the forest code developed on PCs with MPI runs on 
the parallel supercomputers as well.  The performance analyses such as the 
relative error distribution, the timing analysis, and test calculations are 
given in \S \ref{frst:sect:test}.  In \S \ref{frst:sect:summary} the results of this paper are summarized and future 
applications of the forest code to astrophysical problems are also discussed.

\section{A Tree Method}
In the tree method the accuracy of the force calculation is anti-correlated
with the number of gravitational interactions used to calculate the gravity 
force.  This means that if a large acceleration error is allowed, such number 
is kept small leading to the fast force calculation.  However, choosing a 
better way for the tree construction, the structure of tree data, and the 
opening criterion for defining the clusters included in the interaction list,
we can decrease the number of interactions while keeping the acceleration 
error small.  
In this section we describe these factors in detail for use in the 
subsequent sections.

\subsection{Grouping}
\begin{figure}
\begin{center}
\scalebox{0.7}{\includegraphics{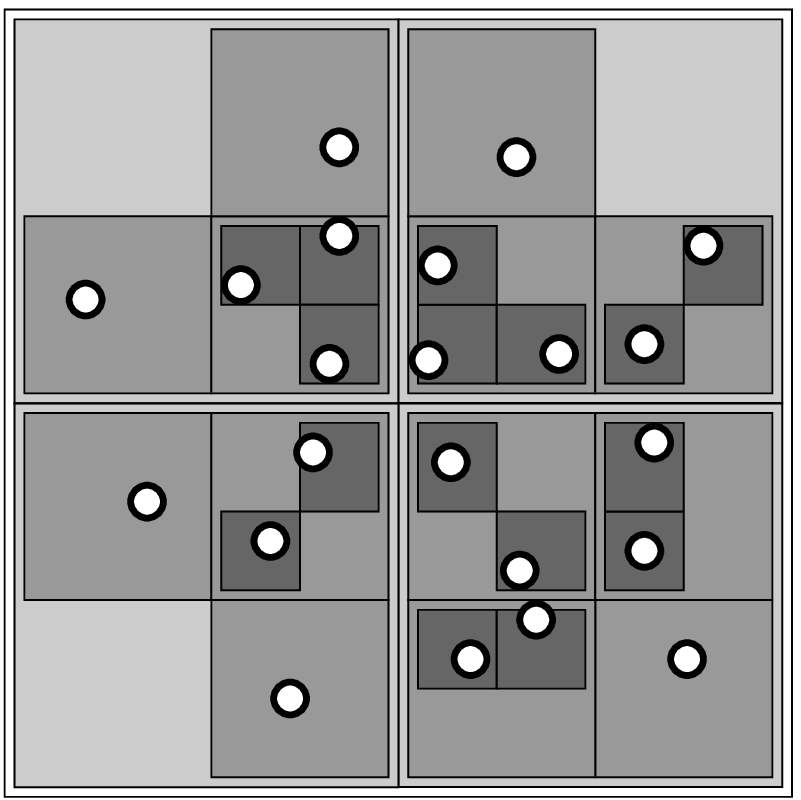}}
\caption[Barnes-Hut tree]{
Tree construction based on the Barnes-Hut (BH) method. 
A top-down scheme is used in constructing an quad-tree in the 
2-dimensional case.  In practice, quad-secting is repeated from the root 
cell which contains all the particles in the system until each of the cells 
contains zero or one particle inside.\label{fig:bhtree}}
\vspace{1cm}
\scalebox{0.7}{\includegraphics{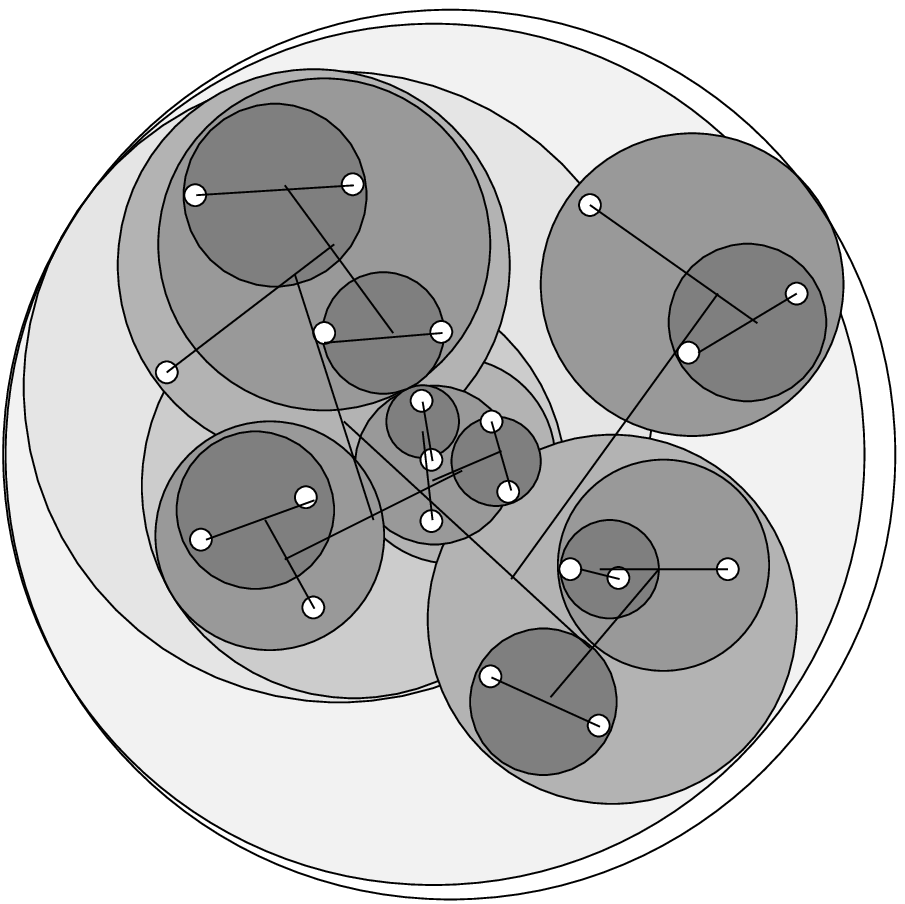}}
\caption[Mutually nearest neighbor tree]{
Tree construction based on the mutually nearest neighbor (MNN) 
method.   A bottom-up scheme is used in constructing a binary tree.   
The mutually nearest neighbors are connected in pairs recursively
until one big cluster eventually
contains all particles and any other clusters in the system.\label{fig:mnntree}
}
\end{center}
\end{figure}

One of two methods proposed for the tree construction is the Barnes-Hut (BH) tree method (Barnes \& Hut 1986).  First of all, a cubic cell which is big enough to contain all the particles is prepared and is called the root cell. Starting from this, the cell is divided into half-sized cubic cells and it is judged how many particles are in each of the divided cells.  If there are more than one particle in the cell, it will be divided further, and this procedure is iterated until there is none or only one particle in each cell.  An oct-tree is then constructed with its leaves regarded as the cells containing one particle inside (Fig. \ref{fig:bhtree}). According to a top-down nature, the BH tree is constructed on the coordinate-dependent algorithm.

Another way to construct the tree is the mutually nearest neighbor (MNN) tree method (Jernigan \& Porter 1989).  In this method, the mutually nearest neighbors are identified as clusters.   Each of clusters  and unclustered particles seeks for another mutually nearest particle and is identified as a new cluster. This procedure is iterated until one big cluster eventually contains all the particles.  In other words, a binary tree is created by connecting them all in pairs (Fig. \ref{fig:mnntree}). This bottom-up nature of the MNN tree is completely opposite to the BH tree, and the algorithm used here corresponds to the coordinate-independent algorithm.

\subsection{Opening criterion\label{OpCrt}}
The clusters are constructed hierarchically using the data arranged in the tree structure as described above. In the tree method, the computational cost is reduced by introducing particle-cluster interactions compared to the PP method. The opening criterion is introduced to judge whether the cluster should be added to the interaction list of the particle. The opening criterion must be set to exclude interactions among near and large clusters which are likely to cause a large acceleration error. A good opening criterion should keep both acceleration error and the number of interactions small.

An opening criterion, originally used in the BH tree method, is given by $ r < L / \theta $, where $r$ is the distance between the particle at which the gravity force is calculated and the center of gravity of the cell to be checked to meet the criterion, and $L$ is the side of the cell.  The opening angle $\theta$ is a user specified parameter to control the resulting acceleration error.  In general, for smaller $\theta$, more cells are opened and smaller errors are attained.  However, if the particle in the cell is localized around the corners of the cell, a large acceleration error is liable to occur (Salmon \& Warren 1994). In order to overcome this difficulty, new opening criteria have been proposed such as the improved BH criterion (Barnes 1994), and the partial absolute multipole acceptance criterion (Salmon \& Warren 1994). The improved BH criterion is given by $r < \delta + L/\theta$, where the distance $\delta$ between the geometrical center of the cell and its center of gravity is included additionally in the original BH criterion. The partial absolute multipole acceptance criterion is given by $\Delta a_{ij} < \Delta a_{max}$, where $\Delta a_{ij}$ is the possible maximum error of the force exerted by $j$-th cell onto the $i$-th particle and $\Delta a_{max}$ is the tolerable partial force error which should be specified by the user.  In the particular case of the monopole criterion, $\Delta a_{ij}$ is expressed analytically as 
\begin{eqnarray}
\Delta a_{ij} = \frac{1}{(r - b_{max})^2} \frac{3 B_2}{r^2},
\end{eqnarray}
where $b_{max}$ is the maximum distance between the positions of the particles in the cell and the center of gravity of the cell, and $B_2$ is the trace of the quadrupole moment tensor (Salmon \& Warren 1994).  The results from different opening criteria will be compared with each other later in \S \ref{frst:sect:erranl}.

\subsection{Vectorization and list sharing}

Unless the opening criterion is met, the cluster of particles is removed from the interaction list, and its daughter clusters and particles are added to the list.  This procedure starting from the root of the cluster tree is iterated recursively, so that the interaction list is constructed in the end.  Additional optimizations for the tree code are possible through the process of constructing this list, by enhancing the efficiency in tree traversal vectorization and list sharing. 

The tree code can easily be written in a recursive way, but such a description can not be vectorized in the stage of tree traversal.  Two special schemes have been developed to vectorize the tree traversal (Hernquist 1990a; Makino 1990).  In particular, Hernquist's scheme is based on the width-first search instead of the depth-first recursive search.

Moreover, since the interaction list is similar to that for the neighbor particles, sharing the interaction list decreases the number of tree traversal.  This list sharing is therefore found to decrease not only the computational time but also the acceleration error of the particles (Barnes 1990).

\subsection{Coordinate dependence of various methods used in the tree 
coding.}
A tree code is made of two parts i.e. tree construction and the force calculation. Moreover, if paralleled, an additional code for the domain decomposition is necessary.  As shown in Table \ref{class}, each of these three is classified as having either the coordinate-dependent algorithm or the coordinate-independent algorithm. The force calculation based on the tree method is coordinate-independent by definition. The BH tree is coordinate-dependent  because the cubic cells are used whose sides are parallel to the axes of the coordinates, while the MNN tree is coordinate-independent.

For a parallel tree code, the ORB method is coordinate-dependent because the space is decomposed into rectangular parallelepipeds by bisecting regions perpendicular to an axis which is cyclically changed. In the HOT method the multidimensional position data for the particles are mapped onto one dimensional integer array or the key array. Then the space decomposition is equivalent to splitting the key array. Since the known mapping methods are coordinate-dependent, the HOT method is also a coordinate-dependent tree parallelization method. On the other hand, the forest method, which will be described in the next section, uses the Voronoi decomposition scheme which is coordinate-independent. 

\begin{table}
\caption[Coordinate dependence of various methods used in the tree coding]{}
\begin{tabular}{lcc}
\multicolumn{3}{c}{Coordinate dependence of various methods used in the tree coding}\\
\hline\hline
		& \multicolumn{2}{c}{Method}\\
\cline{2-3}
Procedure	& Coordinate-dependent	& Coordinate-independent\\
\hline
Force calculation	&PM		&Tree 		\\
Tree construction	&BH tree	&MNN tree	\\
Domain decomposition	&ORB, HOT 	&Forest		\\
\hline
\end{tabular}
\label{class}
\end{table}

\section{The Forest Method -- A New Parallel Tree Method\label{frst:sect:frst}}

In order to realize a fast parallel computing, the data should be divided in such a way that the data communication among different processors is kept small and that the computational time of a certain processor is equal to that for all others.  In the forest method these conditions are simultaneously fulfilled by using the sectional Voronoi tessellation (SVT) and the over-load diffusion (OLD).  The pseudo-code is given in the appendix.

\subsection{Sectional Voronoi tessellation (SVT)}
\begin{figure}
\begin{center}
\scalebox{0.5}{\includegraphics{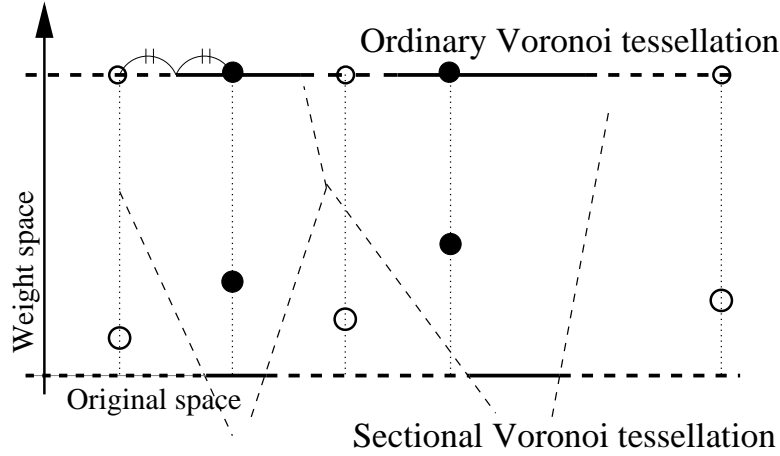}}
\caption[Sectional Voronoi tessellation]{
Illustrative sketch comparing the ordinary Voronoi tessellation 
(OVT) with the sectional Voronoi tessellation (SVT) in the 1-dimensional 
case.  The OVT places a domain border at the equi-distant point between 
the adjacent particles.  However, the SVT changes the domain size according 
to the generator's weight, so that the SVT domain having a higher weight is 
systematically smaller than the OVT domain.
\label{SVT}}
\end{center}
\end{figure}

Suppose that $n$ points or generators $\mbox{\boldmath $x$}_i$ for $i=1,...,n$ are distributed in the $d$-dimensional space $V$, and each point in the space is associated with the nearest generator.  Then the whole space $V$ is divided into $n$ regions of $V_i$, that is,
\begin{equation}
V=\bigcup_{i = 1}^n V_i \;\;\;,
\end{equation}
with the boundary $\partial V_i$ given by
\begin{equation}
\partial V_i = \bigcup_{j = 1}^n (V_i \cap V_j) \;\;\;.
\end{equation}
In the case of the ordinary Voronoi tessellation (OVT), the $i$-th region $V_i$ is the $d$-dimensional polyhedra and the border $V_i \cap V_j$ is the $(d-1)$-dimensional polygon defined as
\begin{equation}
V_i=\{\hspace{0.2cm}\mbox{\boldmath $x$} \hspace{0.2 cm} | \hspace{0.2cm} 
\mbox{\boldmath $x$} \in V,
||\mbox{\boldmath $x$} - \mbox{\boldmath $x$}_i|| \le 
||\mbox{\boldmath $x$} - \mbox{\boldmath $x$}_j||, \forall j \in [1,n]\} 
\;\;\;,
\end{equation}
and
\begin{equation}
V_i \cap V_j =\{\hspace{0.2cm}\mbox{\boldmath $x$} \hspace{0.2 cm} | \hspace{0.2cm} 
\mbox{\boldmath $x$} \in V_i \cup V_j,
||\mbox{\boldmath $x$} - \mbox{\boldmath $x$}_i|| = 
||\mbox{\boldmath $x$} - \mbox{\boldmath $x$}_j||\} \;\;\;,
\end{equation}
respectively, where $||\mbox{\boldmath $x$}||$ represents the Euclidean norm of a vector $\mbox{\boldmath $x$}$ (e.g. Okabe et al. 1992). According to the OVT, the decomposed regions are mutually disjoint, i.e. the intersection of any two decomposed regions has zero-measure in the $d$-dimensional space.

The SVT used in the forest method is one of the generalizations of the OVT. In the SVT each generator is located at ($\mbox{\boldmath $x$}_i, w_i$) in the $(d+1)$-dimensional space, where $w_i$ represents a certain weight to be assigned to the $i$-th generator $\mbox{\boldmath $x$}_i$ in the original $d$-dimensional space.  Accordingly the region $V_i$ and the border $V_i \cap V_j$ in the SVT are defined as
\begin{equation}
V_i=\{\hspace{0.2cm}\mbox{\boldmath $x$} \hspace{0.2 cm} | \hspace{0.2cm} 
\mbox{\boldmath $x$} \in V,
||\mbox{\boldmath $x$} - \mbox{\boldmath $x$}_i||^2+w_i^2 \le 
||\mbox{\boldmath $x$} - \mbox{\boldmath $x$}_j||^2+w_j^2, \forall j 
\in [1,n]\} \;\;\;,
\end{equation}
and
\begin{equation}
V_i \cap V_j =\{\hspace{0.2cm}\mbox{\boldmath $x$} \hspace{0.2 cm} | \hspace{0.2cm} 
\mbox{\boldmath $x$} \in V_i \cup V_j,
||\mbox{\boldmath $x$} - \mbox{\boldmath $x$}_i||^2+w_i^2 = 
||\mbox{\boldmath $x$} - \mbox{\boldmath $x$}_j||^2+w_j^2 \} \;\;\;.
\end{equation}

Following these definitions in the SVT, the weight is used to move the border of the domain.  In other words, the volume of the domain can be changed by changing its weight.  Let $r_{ij}$ be the distance between the $i$-th and the $j$-th generators which are geometrically next to each other. Then the displacement of the border  $V_i \cap V_j$ in the SVT from that in the OVT is calculated as
\begin{equation}
\Delta r_{ij} = r_{ij} \left( \sqrt{1+(w_i^2 - w_j^2)/r_{ij}^2} - 1\right)
\;\;\;.
\label{drij}
\end{equation}
It is evident that if the weights $w_i$ and $w_j$ satisfy $w_i^2 - w_j^2 > \frac{5}{4} r_{ij}^2$, the region $V_i$ does not contain its generator in it.  However, this hardly happens in the forest method where the generators are moved to follow the center of gravity in each step in the domain.  We however note that this, if it happens at all, renders no effects in actual calculations in the forest method.     

\subsection{Over-load diffusion (OLD)}
\begin{figure}
\begin{center}
\scalebox{0.6}{\includegraphics{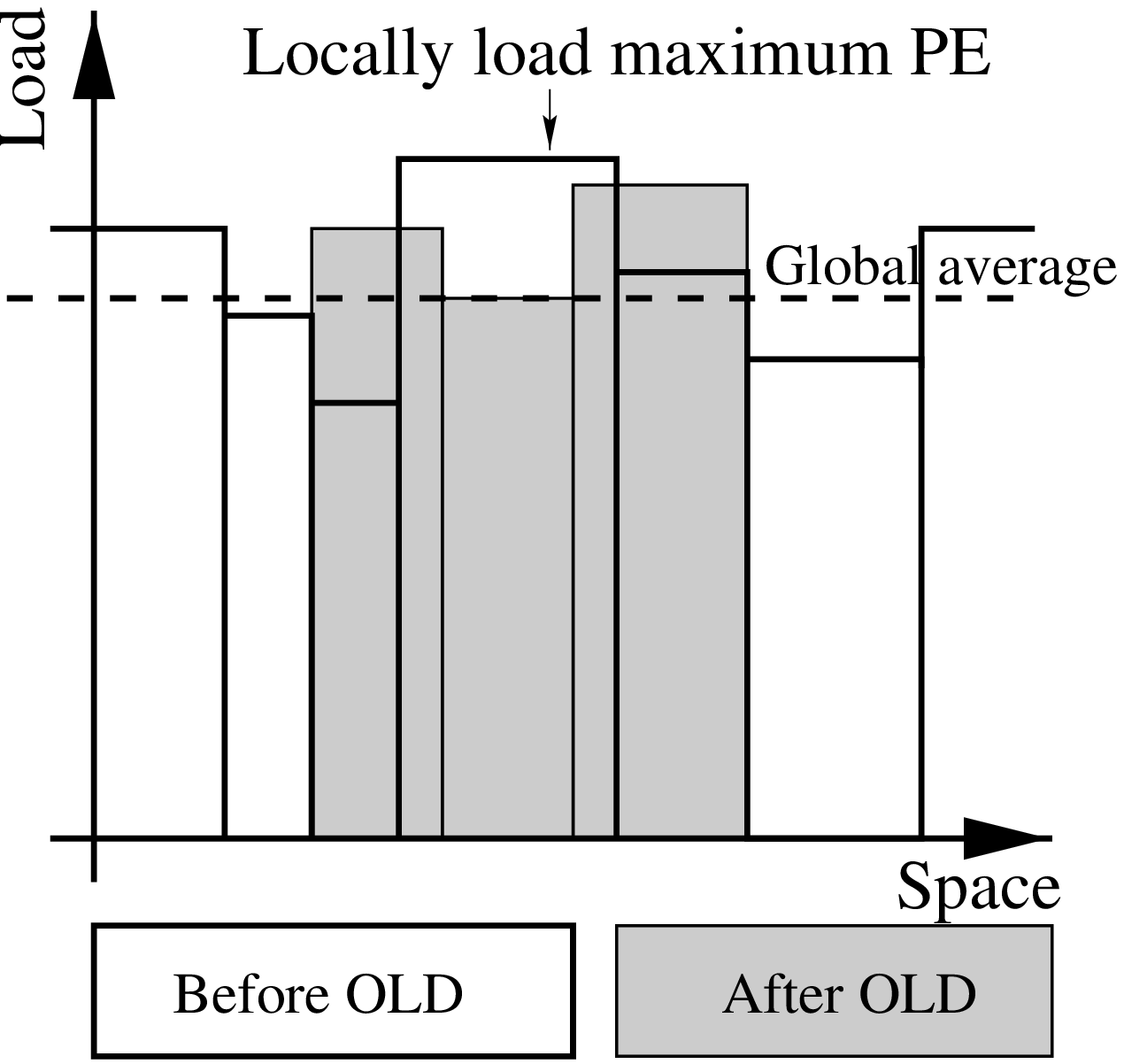}}
\caption[Over-load diffusion: weight tuning]{
Concept of the over-load diffusion (OLD).  The processor 
having a load of local maximum changes its weight in such a way that the 
load is lowered to the global average level.  Accordingly in the next step 
the more load-balanced state is achieved.
\label{old1}}
\vspace{1cm}
\scalebox{0.6}{\includegraphics{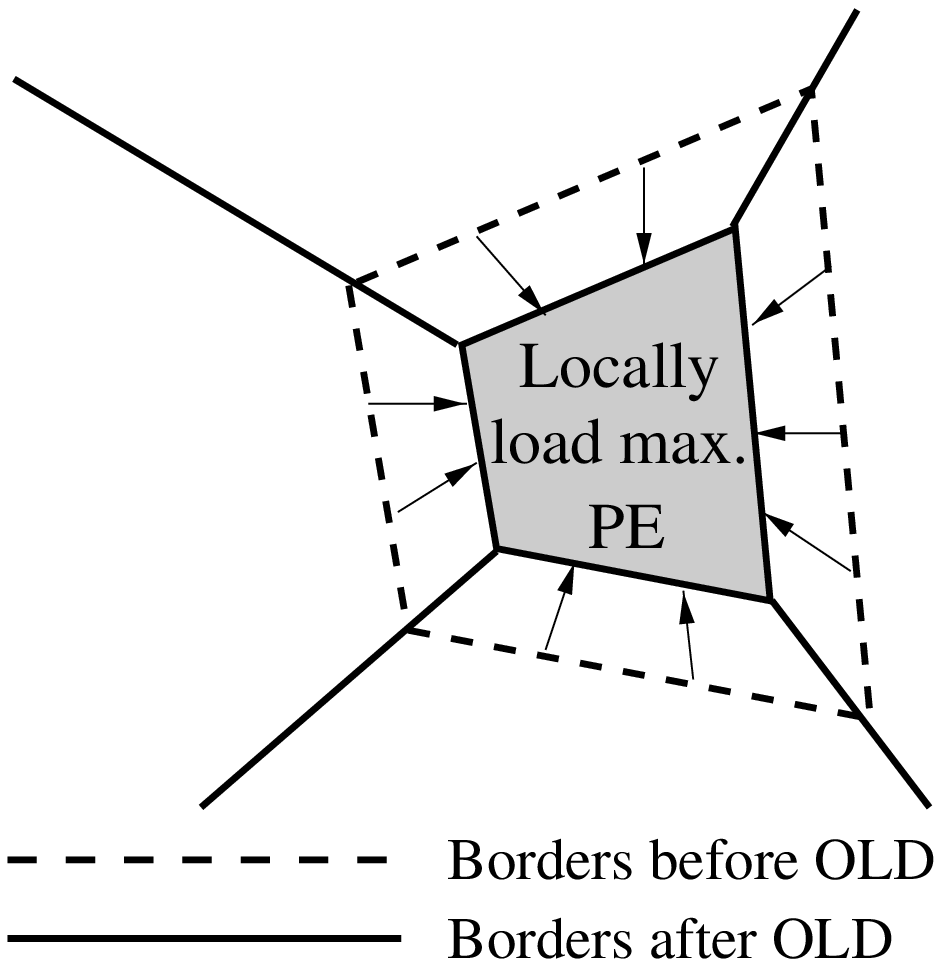}}
\caption[Over-load diffusion: domain contraction]{
Consequence of the over-load diffusion (OLD).  The processor having 
a load of local maximum diminishes its volume according to its weight set 
by the OLD.
\label{old2}}
\end{center}
\end{figure}

The adjustable parameters in the SVT are the positions and weights of the generators only.  Therefore, the loads of the processors or the total number of particles for all the processors cannot be specified directly. However the numbers of particles of a subset of processors which are not geometrically next to one another can be set directly as described in the pseudo-code in the appendix.

In the parallel computing, since the slowest processor determines the speed of the whole calculation, we choose the locally load-maximum processors, each of which has no ambient processor with even higher load, and set their loads equal to the global average load.  This enhances the speed of the locally slowest processors and therefore the speed of the whole calculation as well. If there are some processors having particularly heavy loads initially, such a weight-tuning procedure makes the overload propagate further out to ambient processors as the calculation advances until a load-balancing state is more or less obtained (see \S \ref{SVTandOLD}).  Because of this property of the procedure, we call it the over-load diffusion (OLD).

Another adjustable parameter is the positions of the generators. In the forest method, before starting the calculation, the weights of the processors are initially tuned by shifting generators' positions and applying the SVT and the OLD iteratively such that the particles are distributed equally among the processors. Subsequently in each step of the calculation, the weights are tuned to achieve a load-balancing state by the same methods as the above.

There are many ways to choose the initial positions of generators. For example, each generator can be located at the position of a randomly chosen particle, at the center of gravity of the particles which are randomly assigned to the processors, etc. In any case, the generators are moved toward their respective centers of gravity for particles. As a consequence, the centers of gravity are shifted as if they receive the repulsive forces from others (see e.g. Okabe et al. 1992), Iterating the above generator movement with the OLD weight tuning, domains approach to an equilibrium configuration. Though this configuration has a weak dependence on the initial positions of generators, it does not depend on the coordinate system. Thus, not depending on the initial positions of generators chosen, the forest method is a parallel tree method which is coordinate-independent.

\section{Performance of the Forest Code\label{frst:sect:test}}
We implement our tree code based on the forest method.  This code with the Message Passing Interface is tested on vector parallel processors VPP\-300\-/16R.

\subsection{SVT and OLD\label{SVTandOLD}}
\begin{figure}
\begin{center}
\scalebox{0.6}{\includegraphics{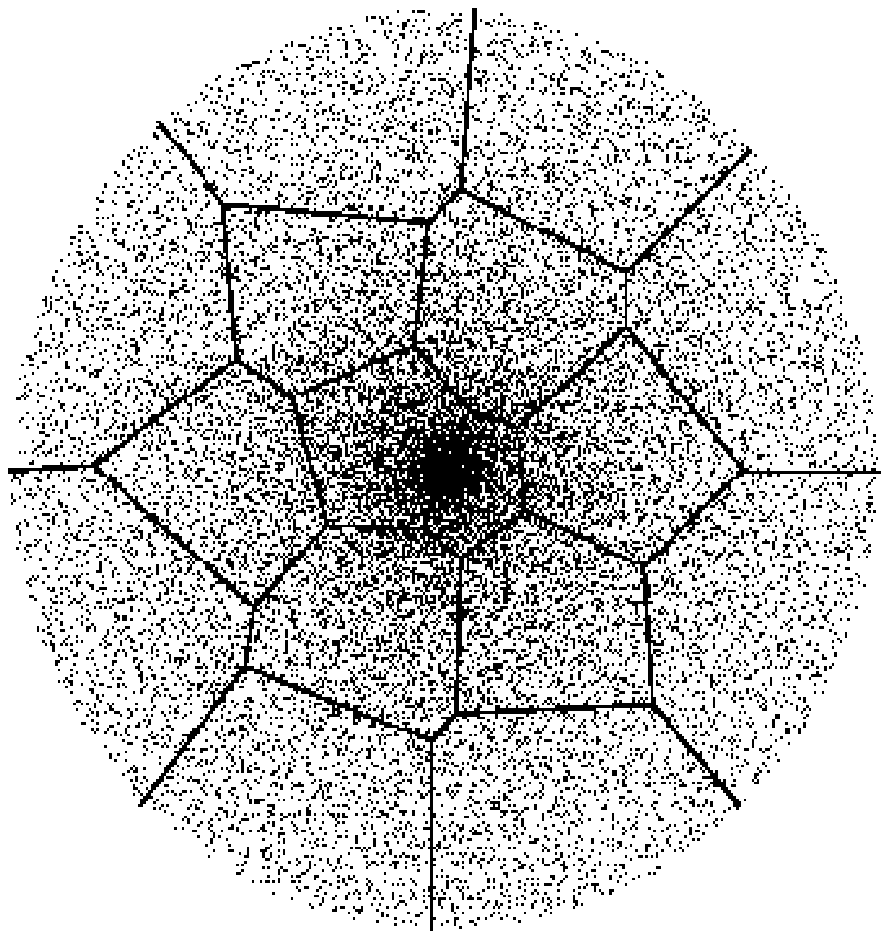}}
\caption[Decomposition example without OLD]{
Sectional Voronoi tessellation (SVT) applied to the truncated 
Mestel disk without weighting by the over-load diffusion (OLD).  
The Mestel disk has an $r^{-1}$ cusp in the surface mass density.   
Hence, although inner domains are apparently smaller than outer domains, 
the smallest domain at the disk center has about three times as many 
particles as the average.
\label{outold}}
\vspace{1cm}
\scalebox{0.6}{\includegraphics{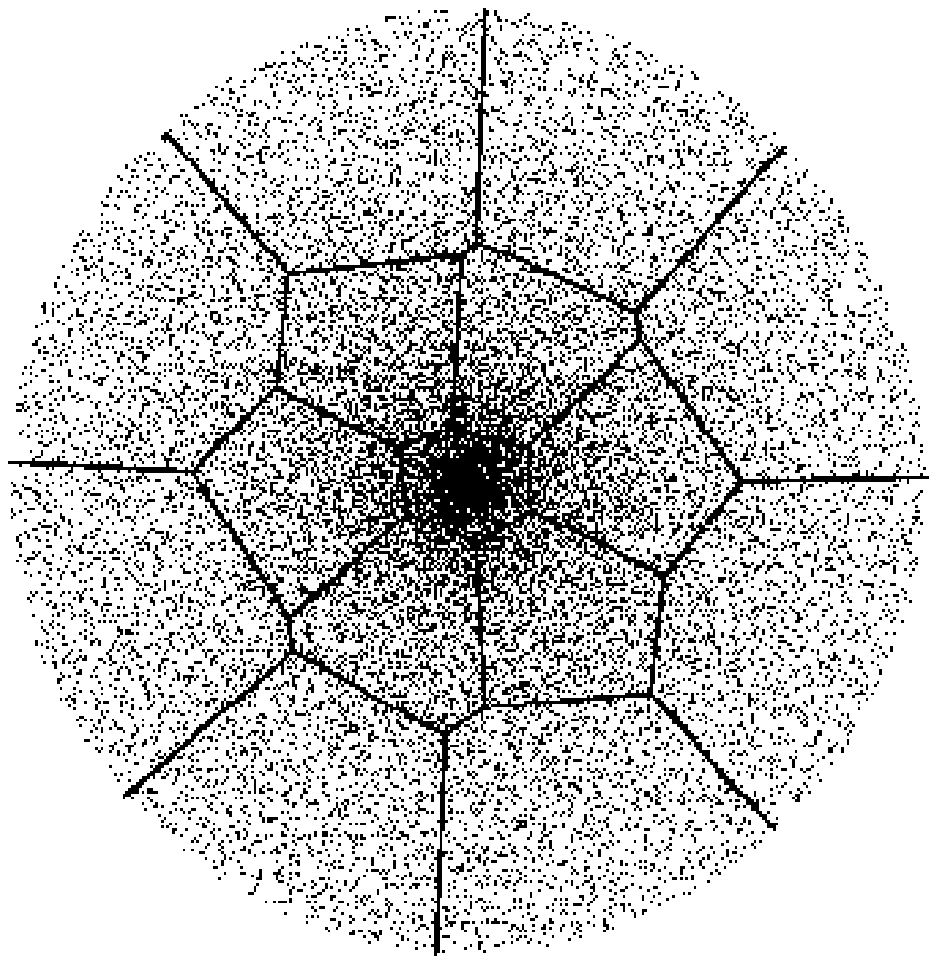}}
\caption[Decomposition example with OLD]{
Sectional Voronoi tessellation (SVT) applied to the truncated Mestel disk.
Same as in Fig. \ref{outold} except that the weighting by the OLD is turned 
on.
The domains at the disk center are smaller than those in Fig. \ref{outold}
without the OLD.   
A symmetrical decomposition with respect to the particle distribution 
is clearly seen.
\label{withold}}
\end{center}
\end{figure}

\begin{figure}
\begin{center}
\scalebox{0.42}{\includegraphics{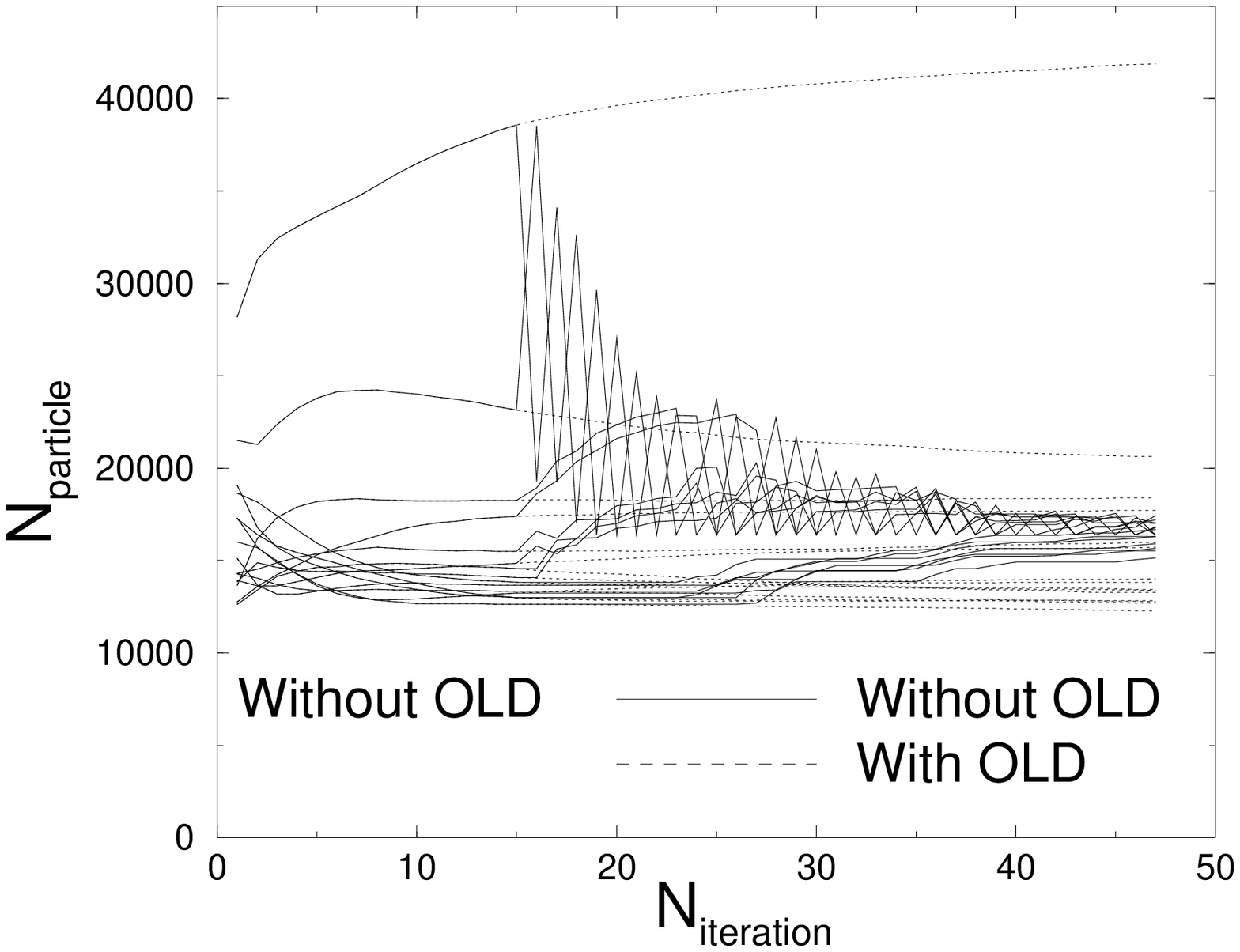}}
\caption[Transition of particle balance]{
The number of particles in each domain plotted against 
the number of steps in the iteration.  The OLD is turned off in 
the first 16 steps and 
the artificial imbalance in the particle number is produced in the domains.  
After the OLD is turned on from the 17th step, the imbalance becomes less
significant by transferring inner particles outwards.
\label{iter}}
\end{center}
\end{figure}

We test the effect of the OLD by choosing a truncated Mestel disk which has an $r^{-1}$ singular surface density cusp at the disk center.  The Mestel disk is suited for the test for the case of the highly inhomogeneous configuration.  Moreover, since the Mestel disk is confined on a plane, the decomposition is easily visualized.

We apply the SVT domain decomposition and move the position of the generator to the center of gravity for the particles in each domain, and this procedure is repeated.  First, the Mestel disk is decomposed without weighting by the OLD (Fig. \ref{outold}). It is apparent from this figure that the disk is decomposed successfully.  However, the central domain has about three times as many particles as the average (Fig. \ref{iter}).  Next, in order to see the effect of the OLD, we turn off the OLD for the first 16 iterations and make artificial load imbalance.  Then we turn on the OLD afterwards from the 17th iteration, so that the domain with the heaviest over-load at the disk center transfers a part of its particles to the other domain at the disk center (Fig. \ref{withold}).  The numbers of particles in the central two domains oscillate transferring their particles to the middle six domains.  Then, the numbers of particles in these inner eight domains oscillate transferring their particles to the outermost eight domains.  In this way the numbers of particles in all domains become equal to each other.

\subsection{Error analysis\label{frst:sect:erranl}}
We investigate the distribution of relative errors and the angular correlation between acceleration and error, based on the improved BH criterion (Barnes 1994; also see \S \ref{OpCrt}) and the partial absolute monopole acceptance criterion (Salmon \& Warren 1994; hereafter SW criterion), using a homogeneous random sphere and Hernquist's (1990b) mass model. In this paper, we define the relative error $|\delta a|/|a|$ as
\begin{equation}
\frac{|\delta a|}{|a|} =  \\
\frac{| \mbox{\boldmath \mbox{\boldmath $a$}}_{PP} - 
\mbox{\boldmath $a$}_{forest}|}{|\mbox{\boldmath $a$}_{PP}|},
\end{equation}
where $\mbox{\boldmath $a$}_{PP}$ and $\mbox{\boldmath $a$}_{forest}$ represent the accelerations of the PP method and the forest method, respectively.  In the SW criterion we use only the monopole term as originally proposed, but the gravitational force is calculated including the quadrupole term.  We use $G = 1$ in the calculations below. The radius and the total mass are given respectively by $R=1$ and $M = 2/9$ for the homogeneous sphere, whereas $R=1$ and $M=1$ for the Hernquist model.

\begin{figure}
\begin{center}
\scalebox{0.4}{\includegraphics{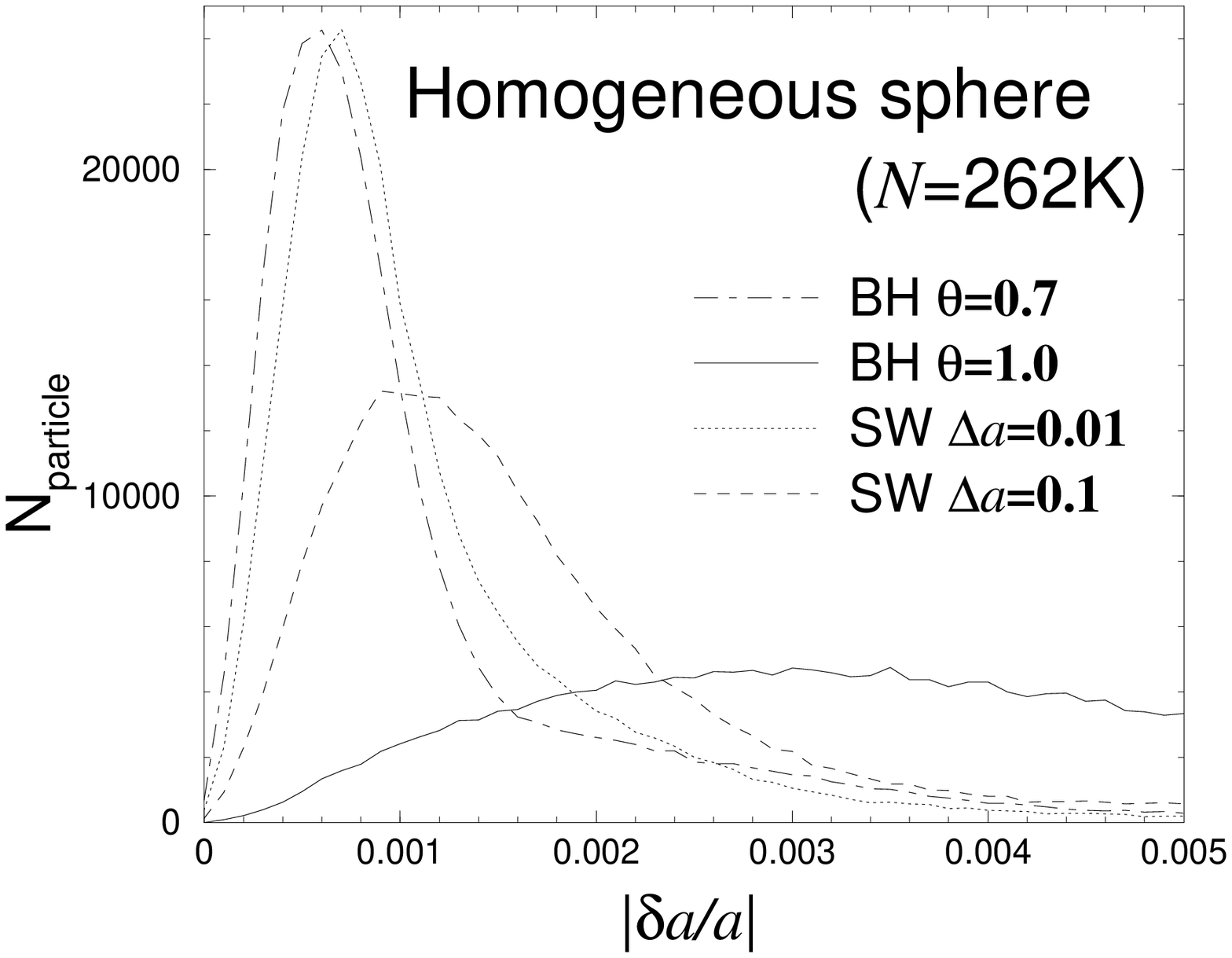}}
\caption[Relative acceleration error for a homogeneous sphere]{
Frequency distribution of relative acceleration error for
a homogeneous random sphere model.
The improved Barnes-Hut (BH) criterion and the
Salmon-Warren (SW) criterion for the opening criterion are separately
used in the analysis. 
The distribution is more extended than in the case of the Hernquist model
(Fig. \ref{hrqerr}),
and this trend is independent of which opening criterion is used in the
analysis.  Especially the improved BH criterion with $\theta = 1.0$ results
in the remarkable change of the distribution compared to that for the 
Hernquist model.  In this particular case the particles having the relative 
error greater than 0.01 comprise 8.5\% in number. \label{hmgerr}}
\vspace{1cm}
\scalebox{0.4}{\includegraphics{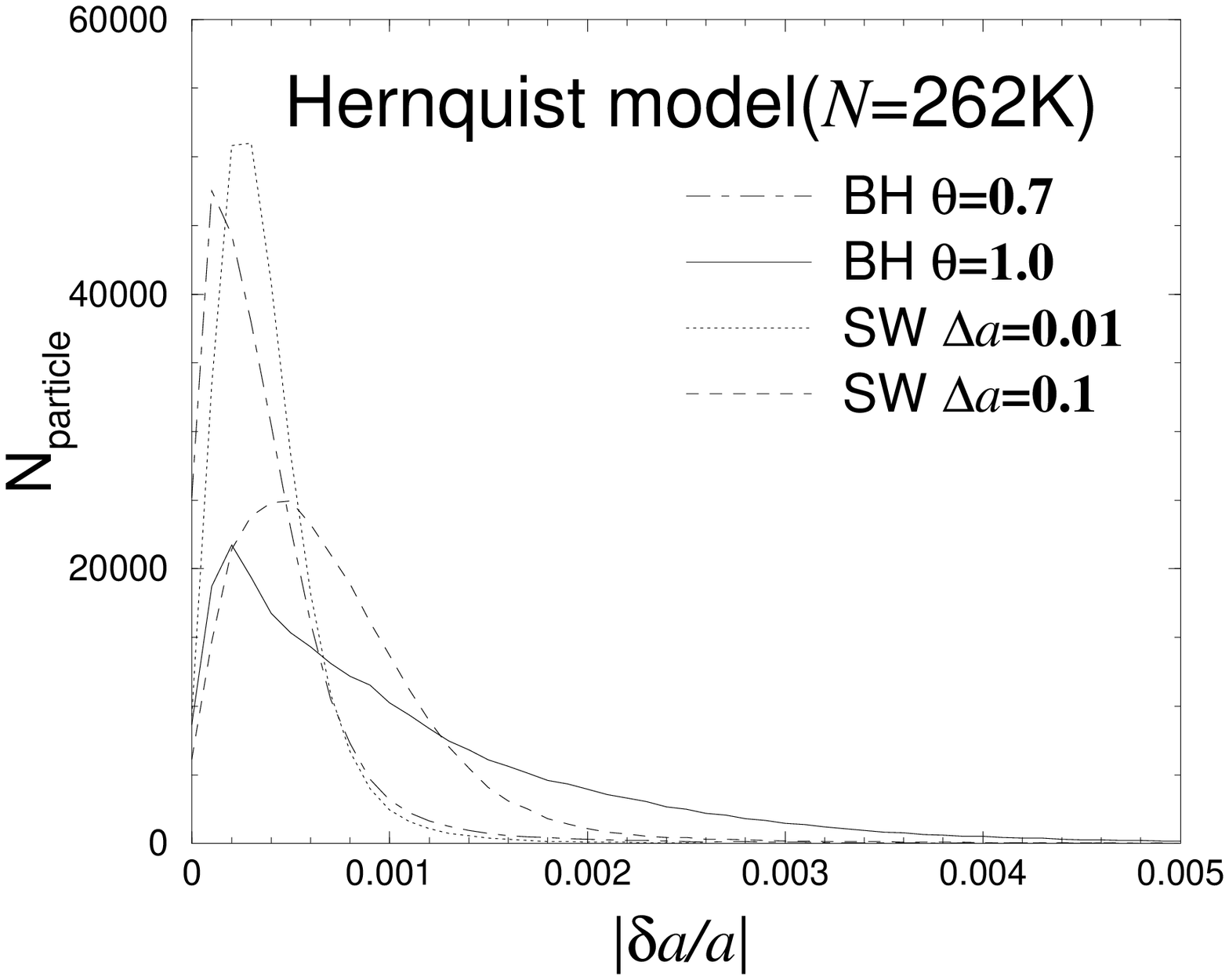}}
\caption[Relative acceleration error for a Hernquist model]{
Frequency distribution of relative acceleration error.  Same as
in Fig. \ref{hmgerr} except for the case of the Hernquist mass model.
The result from the BH criterion is similar to 
Dubinski's (1996) result based on his use of the model galaxy and the model
cluster whose haloes are represented by the Hernquist model.
In the case of the improved BH criterion with $\theta = 1.0$,
the particles having the relative error greater than 0.005 comprise
only 1\% in number. \label{hrqerr}}
\end{center}
\end{figure}

Relative errors from the Hernquist model are systematically smaller than those from the homogeneous sphere, and this tendency is virtually independent of which opening criterion is used in the analysis (Fig. \ref{hmgerr}, \ref{hrqerr}).  However, for the particular case of the improved BH criterion with $\theta = 1.0$, the relative errors are remarkably sensitive to our use of either the homogeneous sphere or the Hernquist model. For example, the particles having the relative errors greater than 0.005 comprise only 1\% in the Hernquist model, whereas those with the errors beyond 0.01 comprise 8.5\% in the homogeneous sphere.
Since the simulation of galaxy formation starting from a homogeneous configuration ends up with a highly clustered configuration, the above result indicates that it is necessary to control the error tolerance parameter while the simulation is in progress.

The relation between the distribution of relative errors and the number of gravitational interactions is shown in Table \ref{tabhmg} for the homogeneous sphere and in Table \ref{tabhrq} for the Hernquist's model. It is apparent from these tables and Figs. \ref{hmgerr} and \ref{hrqerr} that the error distribution from the SW criterion with $\Delta a = 0.01$ is similar to that from the improved BH criterion with $\theta = 0.7$ for both the homogeneous sphere and the Hernquist model. However, the number of gravitational interactions for the SW criterion with $\Delta a = 0.01$ for the Hernquist model is smaller by about a factor 2 than that for the improved BH criterion with $\theta = 0.7$.  

\begin{table}
\caption[Number of interactions and relative error for the homogeneous random sphere model]{}
\begin{tabular}{lccccc}
\multicolumn{6}{c}{Number of interactions and relative error for the homogeneous random sphere model}\\
\hline\hline
&  \multicolumn{2}{c}{Number of interaction / particle}
& &\multicolumn{2}{c}{Relative error$^a$}\\
\cline{2-3} \cline{5-6}
Tolerance parameter	& Particle-Particle	& Particle-Cell	& &Mean	&Median\\
\hline
\hspace{1cm}$\theta = 1.0 \cdots$&108	&310	& &5.090e-3	&4.000e-3\\
\hspace{1cm}$\theta = 0.7 \cdots$&180	&577	& &1.240e-3	&8.257e-4\\
\hspace{1cm}$\Delta a = 0.1 \cdots$&105	&354	& &2.217e-3	&1.503e-3\\
\hspace{1cm}$\Delta a = 0.01\cdots$&95	&496	& &1.185e-3	&9.219e-4\\
\hline
\vspace{-3mm}\\
\multicolumn{6}{l}{$^a$Relative error is defined as $\frac{| \mbox{\boldmath \mbox{\boldmath $a$}}_{PP} - \mbox{\boldmath $a$}_{forest}|}{|\mbox{\boldmath $a$}_{PP}|}$}
\label{tabhmg}
\end{tabular}

\caption[Number of interactions and relative error for the Hernquist mass model]{}
\begin{tabular}{lccccc}
\multicolumn{6}{c}{Number of interactions and relative error for the Hernquist mass model}\\
\hline\hline
&  \multicolumn{2}{c}{Number of interaction / particle}
& &\multicolumn{2}{c}{Relative error$^a$}\\
\cline{2-3} \cline{5-6}
Tolerance parameter	& Particle-Particle	& Particle-Cell	& &Mean	&Median\\
\hline
\hspace{1cm}$\theta = 1.0\cdots$&141	&562	& &1.147e-3	&8.245e-4\\
\hspace{1cm}$\theta = 0.7\cdots$&262	&1292	& &4.424e-4	&3.351e-4\\
\hspace{1cm}$\Delta a = 0.1\cdots$&135	&545	& &7.820e-4	&6.643e-4\\
\hspace{1cm}$\Delta a = 0.01\cdots$&112	&692	& &4.258e-4	&3.716e-4\\
\hline
\vspace{-3mm}\\
\multicolumn{6}{l}{$^a$Relative error is defined as $\frac{| \mbox{\boldmath \mbox{\boldmath $a$}}_{PP} - \mbox{\boldmath $a$}_{forest}|}{|\mbox{\boldmath $a$}_{PP}|}$}
\label{tabhrq} 
\end{tabular}
\end{table}

It is noticeable from Tables \ref{tabhmg} and \ref{tabhrq} that the SW criterion with $\Delta a = 0.01$ gives no appreciable difference in the number of gravitational interactions between the homogeneous sphere and the Hernquist model.  This suppresses the increase in the computational time per step.   Therefore, for the simulation of galaxy formation towards a highly clustered configuration, the SW criterion works better than the improved BH criterion.

\begin{figure}
\begin{center}
\scalebox{0.4}{\includegraphics{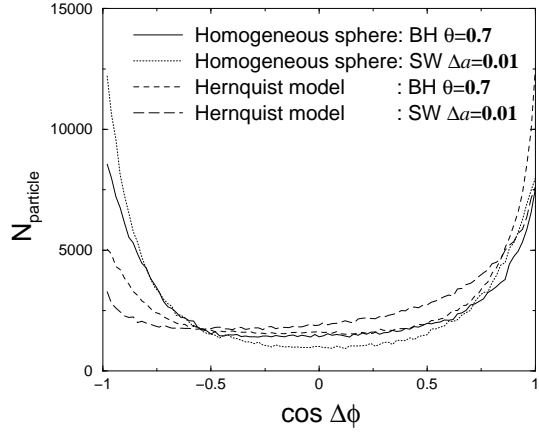}}
\caption[Angular distribution of acceleration error]{
Angular correlation between acceleration error and acceleration 
calculated for various mass models based on the particle-particle (PP) 
method.  The peak at $\cos \Delta \phi =-1 $ is contributed from the 
particles near the center of the system, while the peak at 
$\cos \Delta \phi=+1$ is from the particles near the surface of the
\label{cosN}}
\end{center}
\end{figure}

We have examined not only the error size but also the direction of the error distribution (Fig. \ref{cosN}).  If the acceleration error vectors are distributed isotropically, the lines must be horizontal in Fig. \ref{cosN}, but there are obviously two peaks at $\cos\Delta\phi =-1$ and $+1$.  For the homogeneous sphere, the peak at $\cos\Delta\phi = -1$ is contributed from the particles near the center and the peak at $\cos\Delta\phi =+1$ from the particles near the surface.  The angular correlation between acceleration and error slightly depends on the opening criterion, but its dependence is weaker than that on the spatial distribution of the particles.

\subsection{Timing analysis}
\begin{figure}
\begin{center}
\scalebox{0.4}{\includegraphics{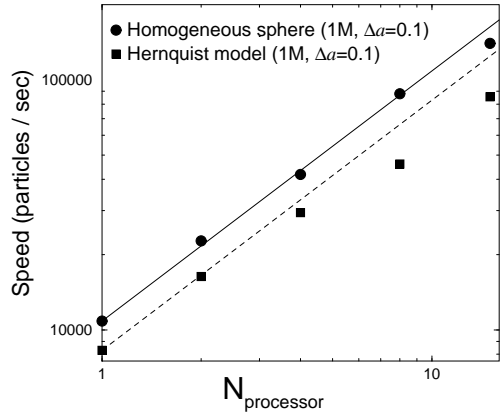}}
\caption[Speed of the forest code]{
The number of particles for which the force is calculated per
second on VPP300/16R using the forest code. The solid line indicates the 
ideal speed which is estimated from the serial code for the homogeneous 
sphere model, and the dashed line for the Hernquist model.  The SW 
criterion with $\Delta a = 0.1$ is used for both cases.  It is
shown that the ideal speed is achieved in the actual calculation by the 
\label{timana}}
\end{center}
\end{figure}

The speed of the forest code is tested on the Fujitsu VPP300/16R which is a system of vector parallel processors.  The SW criterion with $\Delta a = 0.1 $ is used and the number of particles is set to be $1.0 \times 10^6$.  Both the homogeneous sphere and the Hernquist model are considered in the analysis.  

The ratio of the speed for the homogeneous sphere relative to that for the Hernquist model is independent of the number of processors.  The speed of the code is almost the same as the ideal speed estimated from the serial code.  

The load balance is also tested.  Let $l$ be the ratio of the mean elapsed time to the maximum elapsed time over the processors:
\begin{eqnarray}
l = \frac{\left(\sum\limits_{i = 1}^{N_{processor}} t_{elp, i}\right) / N_{processor}}{\max(t_{elp, 
1}, \ldots, t_{elp, N_{processor}})},
\end{eqnarray}
where $t_{elp, i}$ is the elapsed time of the $i$-th processor. Then this quantity measures how evenly the loads are distributed over the processors.  We note that $l$ converges to unity as the simulation goes on with increasing the number of time steps.

\subsection{Tests}
\begin{figure}
\begin{center}
\scalebox{0.4}{\includegraphics{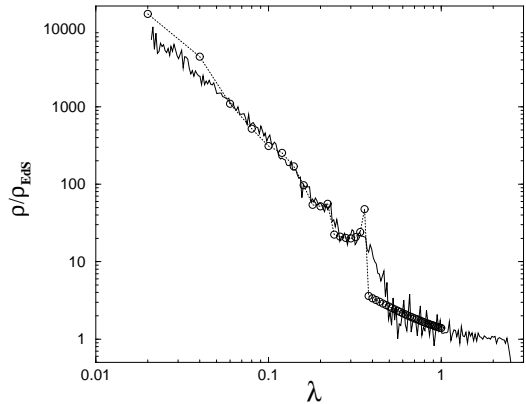}}
\caption[Secondary infall test]{
Radial density profile of the cosmological secondary infall.
The circles connected by the dotted line represent the 
self-similar solution taken from Table 4 of Bertschinger (1985).
Density is normalized to unity by the background density, and
$\lambda$ represents the radius normalized by the turnaround radius.
The particles just inside the outermost region are beginning 
to infall towards the center.
The particles, which have just crossed the shell and are in the stage of 
the secondary turnaround, are viewed as the first caustic. 
The density profile in the innermost region approaches $r^{-9/4}$.  
\label{berttest}}
\end{center}
\end{figure}

The acceleration error from the forest code has been analyzed in the preceding sections, but the acceleration error, coupled with time integration, causes the deviation of particle trajectories. Therefore, the comparison of the $N$-body result with an analytic or a semi-analytic solution for some ideal problem is important.

We choose the self-similar solution for cosmological secondary infall in the Einstein-de Sitter universe (Bertschinger 1985), because it is very similar to the formation of a virialized object of our concern. We perform a simulation with $10^5$ particles, starting from an expanding homogeneous sphere with escape velocity. We have added a particle at the center of the sphere which has a 1/20 of the total mass of the system. The density profile reaches an asymptotic form as shown by the solid line in Fig. \ref{berttest}. The particles in the outermost region are on the Hubble flow and their density is equal to the background density of the universe.  The particles just inside the outermost region are beginning to infall towards the center. The particles, which have just crossed the shell and are in the stage of the secondary turnaround, are viewed as the first caustic. However, the first caustic represented in the test simulation is blunt compared to the semi-analytic solution. The density profile in the innermost region approaches $r^{-9/4}$ which is given by the analytic solution of the secondary infall.  This comparison clearly indicates that the forest code gives a reliable simulation result.

\section{Summary\label{frst:sect:summary}}
We have developed a new parallel tree method in which the sectional Voronoi tessellation and the over-load diffusion are incorporated.  Since the tree construction scheme in this method is forest-like, we called it the forest method.  

A code based on the forest method is developed on our own PCs to run on supercomputers with the Message-Passing Interface (MPI). The performance of the code is checked on VPP300/16R which is a system of vector parallel processors, and it is confirmed that the speed of the code increases almost linearly with increasing the number of processors.

The correlation between the distribution of relative acceleration errors and the number of interactions per particle depends not only on the spatial distribution of particles but also on the opening criterion used. The Salmon-Warren criterion is found to be more efficient than the improved Barnes-Hut criterion, especially when highly clustered configurations are not avoided in simulations.

With this code, we can carry out a very large simulation using as many as $10^7$ or more particles allowing for a wide dynamic range. A number of applications are possible including the study of galaxy formation and large-scale structure in the universe.  The dynamical evolution of galaxies, either isolated in general fields or aggregated in pairs or clusters, is also investigated in detail by using the forest code here developed.

\subsection*{acknowledgements}
We thank Naohito Nakasato for useful discussions.
All calculations have been performed on the VPP\-300\-/16R 
of the National Astronomical Observatory, Japan.  This work has
been supported in part by the Research Fellowship of the Japan Society for
the Promotion of Science for Young Scientists (6867) and the Grant-in-Aid 
for COE research (07CE2002) of the Ministry of Education, Science, and
Culture in Japan.

\section*{References}
\begin{list}{}{\itemindent=-8mm \parsep=-2mm \topsep=0mm }
\item Anderson, C. R., 1992, In SIAM Journal on Scientific and Statistical Computing, 13, 923
\item Appel, A. W., 1985, In SIAM Journal on Scientific and Statistical Computing, 6, 85
\item Barnes, J. \& Hut, P., 1986, Nature, 324, 446
\item Barnes, J., 1990, J. Comp. Phys., 87, 161
\item Barnes, J. E., 1994, In Computational Astrophysics, Eds. J. Barnes et al.
\item Bertschinger, E., 1985, ApJS, 58, 39 
\item Couchman, H. M. P., 1991, ApJ, 368, 23L
\item Dubinski, J., 1996, NewA, 1, 133 
\item Efstathiou, G., Davis, M., Frenk, C., White, S. D. M., 1985, ApJS, 57, 241
\item Greengard, L. \& Rokhlin, V., 1987, J. Comp. Phys., 73, 325
\item Hernquist, L., 1987, ApJS, 64, 715
\item Hernquist, L., 1990a, J. Comp. Phys., 87, 137 
\item Hernquist, L., 1990b, ApJ, 356, 359
\item Hockney, R. W. \&  Eastwood, J. W., 1988, Computer Simulation Using Particles (Bristol: Institute of Physics Publishing)
\item Jernigan, J. G. \& Porter, D. H., 1989, ApJS, 71, 871
\item Makino, J., 1990, J. Comp. Phys., 87, 148
\item Miller, R. H., 1978, ApJ, 223, 122
\item Okabe A., Boots, B. \& Sugihara, K., 1992, Spatial Tessellations Concepts and Applications of Voronoi Diagrams (Chichester: John Wiley \& Sons Ltd)
\item Salmon, J. K., 1990, Ph.D. Thesis, California Institute of Technology
\item Salmon, J. \& Warren, M. S., 1994, J. Comp. Phys., 111, 136   
\item Sugimoto, D., Chikada, Y., Makino, J., Ito, T., Ebisuzaki, T., \& Umemura, M., 1990, Nature, 345, 33 
\item Villumsen, 1989, ApJS, 71, 407
\item Warren, M. S. \& Salmon, J. K., 1993, in Supercomputing '93, 12 (Los Alamitos: IEEE Computer society)
\item Warren, M. S. \& Salmon, J. K., 1995, Computer Physics Communications, 87, 266
\end{list}

\clearpage
\appendix
\section*{Appendix}
\section{The pseudo-code of the OLD with the SVT\label{psdcd:frst}}

In the forest method the OLD operates together with the SVT. The pseudo-code for the main part of this OLD is given as follows:

{\tt
\noindent
Move the generator to the center of gravity of the particles in the domain.\\
\#Begining of the OLD \\
Check whether the processor is locally load-maximum (LLM) or not.\\
If the processor is LLM,\\
\hspace*{15mm}calculate with the signed distances of the particles \\
\hspace*{20mm}to the nearest domain border.\\ 
\hspace*{15mm}sort their signed distances in a descending order.\\
\hspace*{15mm}read the sorted distance in the 
$(\frac{<Load>}{Load_{here}} \times n_{particle}$)-th row.\\
\hspace*{15mm}redefine $w_{here}$ as $w_{here}+$ the above distance.\\
(End of if)\\
\#End of the OLD \\
\#Begining of the SVT\\
\noindent
For each particle,\\
\hspace*{15mm}calculate the distance to the processor to which it belongs.\\
\hspace*{15mm}if this distance is greater than the distance to the nearest
neighbor processor (NNP),\\ 
\hspace*{30mm}send the particle to the NNP.\\
(End of for) \\
\#End of the SVT
}

The distance is always calculated in the weight-added space. The sign of the distance to a plane is defined to be positive, when the particle and the generator lie on the same side against  the plane.
\end{document}